\documentclass[a4paper,11pt]{article}  % Single-column

\usepackage[utf8]{inputenc}
\usepackage[T1]{fontenc}
\usepackage{lmodern}          % Modern Latin font
\usepackage{geometry}         % Page geometry
\usepackage{authblk}          % Author & affiliation formatting
\usepackage{amsmath, amssymb} % Math symbols
\usepackage{graphicx}         % Figures
\graphicspath{{figures/}}
\usepackage{caption}
\usepackage{subcaption}       % For subfigures
\usepackage{tabularx}         % Tables with automatic column width
\usepackage{booktabs}         % Nice table formatting
\usepackage{siunitx}          % SI units
\usepackage{abstract}         % Abstract formatting
\usepackage{listings}         % Code listings
\usepackage{microtype}        % Better line breaking and typography
\usepackage{xurl}             % Line breaks in URLs
\usepackage[hidelinks]{hyperref} % Clickable links
\usepackage[numbers,sort&compress]{natbib} % Bibliography
\usepackage{csquotes}         % Curly quotation marks
\usepackage[section]{placeins}   % Keep floats within sections
\usepackage{dblfloatfix}      % Allows bottom placement of full-width floats

\title{\textbf{The electron double-slit experiment from an ISP perspective}}
\author[1]{David LeBlond}
\affil[1]{Independent Researcher, United States,
\href{mailto:david.leblond@sbcglobal.net}{david.leblond@sbcglobal.net}}
\date{\today}

\begin{document}

	\maketitle
	\begin{onecolabstract}
		This paper presents a pedagogical model, and accompanying R code, of the electron double-slit experiment using the perspective of indivisible stochastic processes. The approach offers 
		an alternative lens on quantum probability and coherence phenomena, emphasizing a statistical rather than purely wave-mechanical interpretation.
	\end{onecolabstract}
	\vspace{1em}
	
	\section{Background and motivation}
	Quantum mechanics (QM) is often presented through an axiomatic paradigm similar to a branch of algebra or geometry. For instance:
	\begin{enumerate}
	  \item The state of a quantum system is described by a vector in a complex Hilbert space. 
	
	  \item Every physical observable corresponds to a Hermitian linear operator.
	
	  \item The only possible measurement outcomes are operator eigenvalues.
	
	  \item The probability of obtaining a specific measurement outcome is determined by the Born rule.
	
	  \item Isolated quantum systems evolve via the time-dependent Schrödinger  unitary transformation.
	
 	  \item The Hilbert space for composite systems is the tensor product of the Hilbert spaces of the individual subsystems.
	\end{enumerate}
	
	These axioms provide impressive predictive power, yet resist deeper underlying mechanistic justification typical of a fertile scientific model. For the student of science accustomed to intuitive physical frameworks, the arcane imagery of wave-particle duality, superposition, entanglement, measurement collapse, and tunneling lack compelling mechanistic inspiration. The search for a deeper understanding is irresistible.
	
	It has recently come to light that many counter-intuitive quantum mechanical phenomena can be understood as indivisible stochastic processes (ISP) leading to a more accommodating indivisible quantum theory (see \cite{lecturenotes,Barandes2023,Barandes2024,Barandes2025a,Barandes2025c,Barandes2025b}). These axioms seem simpler and more physically transparent:
	
	\begin{enumerate}
		\item Kinematics consists of a fixed configuration space of the modeled system.
		
		\item Configurations evolve dynamically over time via transition probabilities that condition on  division events.
		
		\item At a given time, the set of configurations has a standalone probability distribution.
	\end{enumerate}

	Axioms 1 and 3 are based on notions of classical probability. The Born rule is treated as an identity rather than a postulate. Non-Markovian stochastic dynamics can exhibit counter-intuitive (yet understandable) behavior mirroring those of quantum systems. This recognition offers the hope for a connection between the axioms of QM and their underlying physical (or at least statistical) basis.
	
	\section{Purpose of the present work}
	It is instructive to reproduce the predictions of QM via ISP-based modeling. A coarse-grained ($N=2$ positions) illustration of the famous electron double slit experiment (see  \cite{lecturenotes}) was kindly made available. The purpose here is to use the arguments presented in that illustration to reduce the coarse graining (to $N=2000$ positions). The reduction in coarse graining introduces some additional aspects that require computation. However, it more closely models a hypothetical experimental setup and provides a compelling graphical demonstration.
	
	Section 3 outlines the arrangement and assumptions behind a double-slit experiment. Section 4 presents a graphical perspective. An ISP-based model and corresponding predictions are described in Sections 5 and 6, respectively. A brief discussion follows in Section 7. The R code employed is provided in the Appendix. 
	
	\section{Description of the \enquote{experimental} set-up}
\begin{figure}[htbp]
		\centering
		\includegraphics[width=0.9\textwidth]{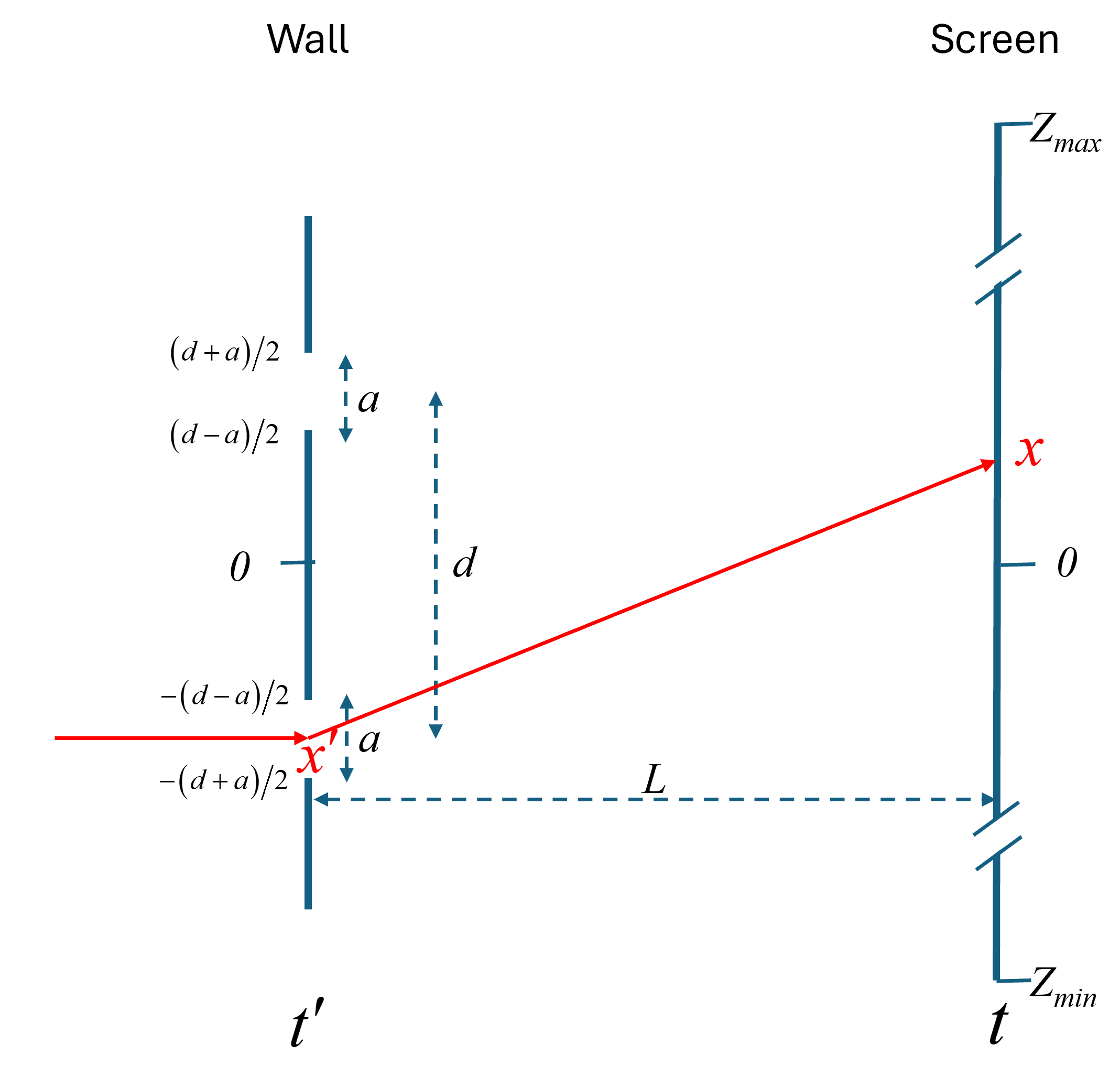}
		\caption{Double-slit experimental setup. Red arrows indicate one possible electron path.}
		\label{fig:interference}
	\end{figure}

	Helpful theoretical and physical background of the double-slit experiment, from a quantum mechanical perspective, are available from \cite{beau_2012}, \cite{cernak_2024}, \cite{jones_2015}, and \cite{rolleigh_2014}. A common experimental setup is diagrammed in Figure~\ref{fig:interference}.
	
	The vertical dimension measures distance from the point midway between the upper and lower slits and this origin is common to both the slit wall and detector screen. However, the distance scale at the detector screen ($x$ at time $t$) is much greater (by a factor of $\approx 4\times 10^7$ fold) than the distance scale at the slit wall screen ($x'$ at time $t' < t$).
	
	The horizontal dimension measures distance between the slit wall and the detector screen ($L=\SI{1}{\metre}$). 
	
	\subsection{Assumptions about the electrons}
	The red arrows in Figure~\ref{fig:interference} indicate one possible path the electron might take. We assume... 
	
	\begin{itemize}
	\item Electrons of mass $m = \SI{9.109e-31}{\kilogram}$ and wavelength $\lambda= \SI{1.23e-10}{\meter}$ approach the wall normal to the surface, one at a time, with velocity $v=h/(\lambda\times m)=\SI{5.914e6}{\meter\per\second}$, where Plank's constant $h=\SI{6.6261e-34}{\joule\cdot\second}$. This can be achieved by acceleration across a potential of approximately $\SI{100}{\volt}$. 
	
	\item The velocity is assumed low enough to ignore relativistic effects. 
	
	\item The electron gun is sufficiently distant from the wall that the electrons approach the wall surface perpendicularly. 
	
	\item A uniform Probability Density Function (PDF) of wall arrival positions within the slits with zero probability density outside the slits (we experimentally ignore any electrons outside the slits at time $t'$). 
	
	\item As described below, the uniform PDF within the slits and the resulting PDF at the screen at time $t$ are approximated (coarse-grained) as discrete Probability Mass Functions (PMFs) with $N=2000$ intervals across the slits and screen, designated by $x'_{i'}$ and $x_i$, respectively, where $i'$ and $i=1,2,...,N$. 
	
	\item The electron state transition between the wall ($x'$ at time $t'$) and screen ($x$ at time $t$) can be described by a path integral \cite{feynman}. 
	
	\item The distance $L=\SI{1}{\meter}$ between the wall and screen is very large relative to the slit width $a=\SI{0.15e-8}{\meter}$ and distance between slits in meters $d=\SI{0.615e-8}{\meter}$ and also large enough relative to the range of interest across the screen ($Z_{max}-Z_{min}=\SI{0.3}{\meter}$) such that dependence of the transition time ($t-t'$) on the electron diffraction angle $\theta=\tan^{-1}((x-x')/L)$ can be ignored.
		\end{itemize}
	
	\subsection{Assumptions about the slit wall}
	
	There are 2 identical slits of width $a=\SI{0.15e-8}{\meter}$ and separation distance of $d=\SI{0.615e-8}{\meter}$. The length of the slits (measured perpendicular to the diagram) are assumed infinite so that diffraction effects in this dimension are ignored.
	
	Each slit is coarse grained into $N/2=1000$ locations $x'_{i'}$, $i'=1,2,...,N$ where $i'\le N/2$ and $i'>N/2$ correspond to the lower and upper slits, respectively.  In fact there will be many more positions at the slit wall but these are not active because any electron not at a slit position is ignored.
	
	We take $x'=0$ as the point midway between the slits so that $x'_{i'}$ is the distance from the point midway between the 2 slits. Thus $-(d+a)/2 \le x' \le -(d-a)/2$ is the range of the lower slit and $(d-a)/2 \le x' \le (d+a)/2$ is the range of the upper slit.
	
	The width of the $N/2$ coarse-grained discrete intervals within each slit is thus $\Delta_{slit} = 2a/N$.  
	
	\subsection{Assumptions about the environmental qubit}
	An environmental 2-state qubit, having values of 1 (upper slit) or 2 (lower slit) indexed by $e'$ (which has a corresponding value), is assumed to operate faithfully at a time infinitesimally smaller than $t'$. The qubit has a default state of 1 but may have 3 different behaviors:
	\begin{enumerate}
		\item Remembers: The qubit state changes from 1 to 2 if the electron is in the lower slit at time $t'$ and is thus a perfect measuring device. The state does not change from time $t'$ to $t$. The qubit state at time $t$ is indexed by $e$ and thus $e=e'$.
		
		\item Forgets: Similar to the Remembers behavior, except $e$ always reverts to 1 regardless of the value of $e'$.
		
		\item None: The qubit always remains in state 1 so that $e'=e=1$. Effectively, the qubit is inactive and does not interact with the electron.
	\end{enumerate}
	
We assume here transition probabilities for the qubit are either 1 or 0 depending on its behavior. However, other transition probabilities could be considered.

\subsection{Assumptions about the screen}
A detector screen is located at $L=\SI{1}{\meter}$ from the slit wall and registers the electron discrete position $x_i$ at the screen at time $t$. There is a one-to-one mapping of slit positions $x'_{i'}$ and screen positions $x_i$. However, $x_i \ne x'_{i'}$. The intensity at the screen is sampled over the range $Z_{min}=-0.15 \le x_i \le +0.15 = Z_{max}$ at intervals of $\Delta_{screen} = (Z_{max}-Z_{min})/N$ which is much greater than the corresponding slit dimensions.

Each position (and corresponding qubit state) at the slit wall will have an exact mirror at the screen. However, only selected screen positions will be 
sampled. Therefore, the screen, like the wall, will also have $N$ positions with 2 qubit states for each position.

\section{Graphical representation of the transition matrix}
\begin{figure}[htbp]
	\centering
	\includegraphics[width=0.9\linewidth]{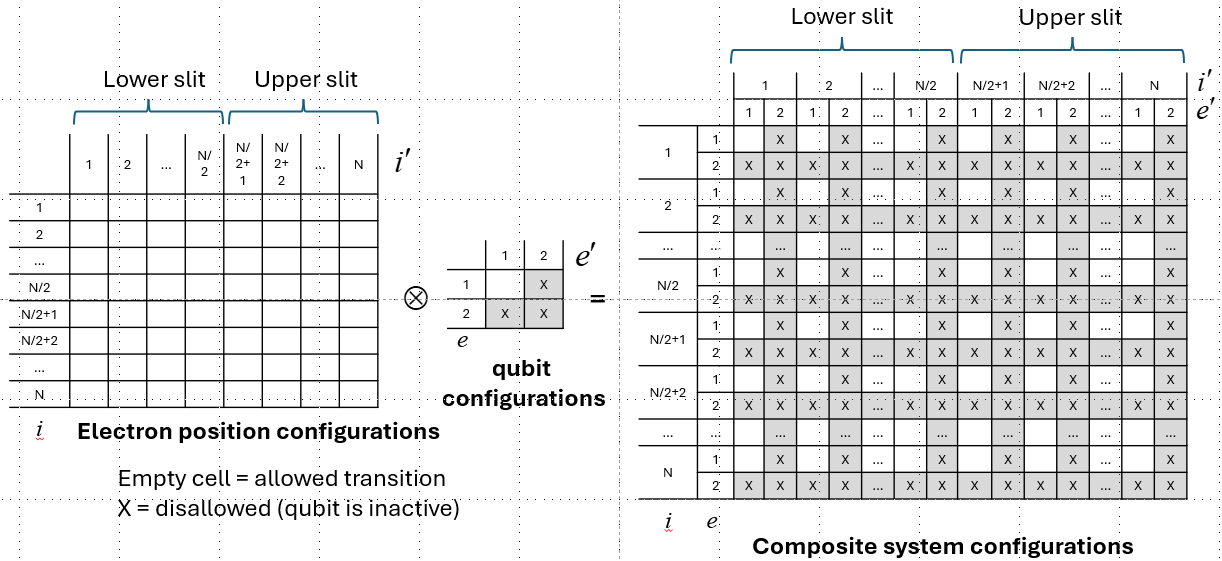}
	\caption{Transition matrix for composite system-qubit configuration states when qubit is inactive (qubit behavior = \enquote{None}).}
	\label{fig:inactive}
\end{figure}

\begin{figure}[htbp]
	\centering
	\includegraphics[width=0.9\textwidth]{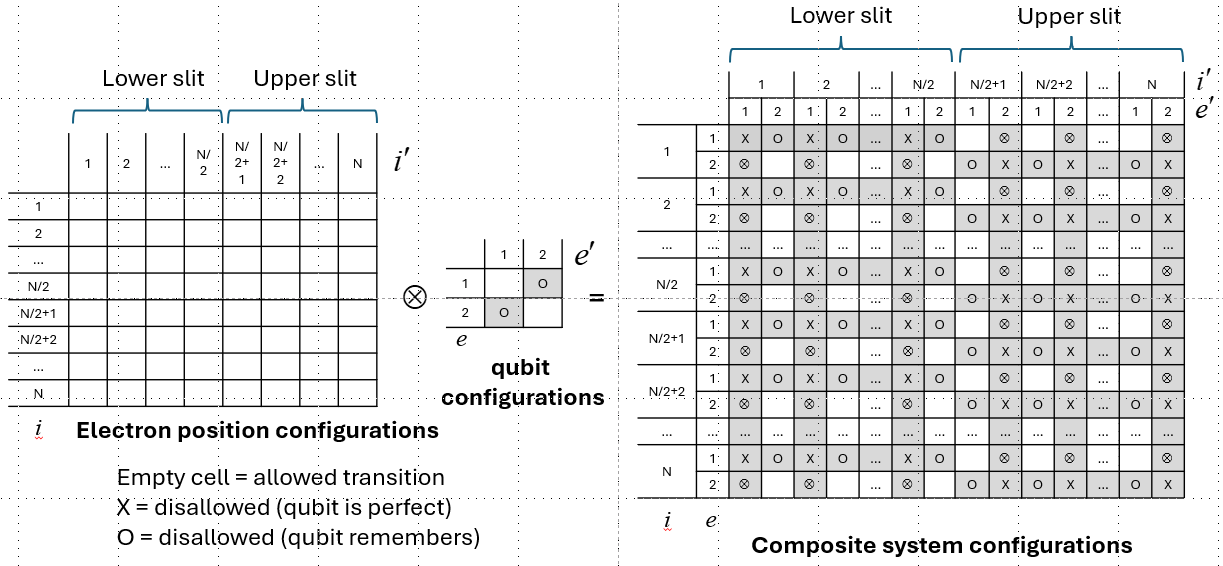}
	\caption{Transition matrix for composite system-qubit configuration states when qubit detects and \enquote{Remembers} the electron's entry slit.}
	\label{fig:remembers}
\end{figure}

\begin{figure}[htbp]
	\centering
	\includegraphics[width=0.9\textwidth]{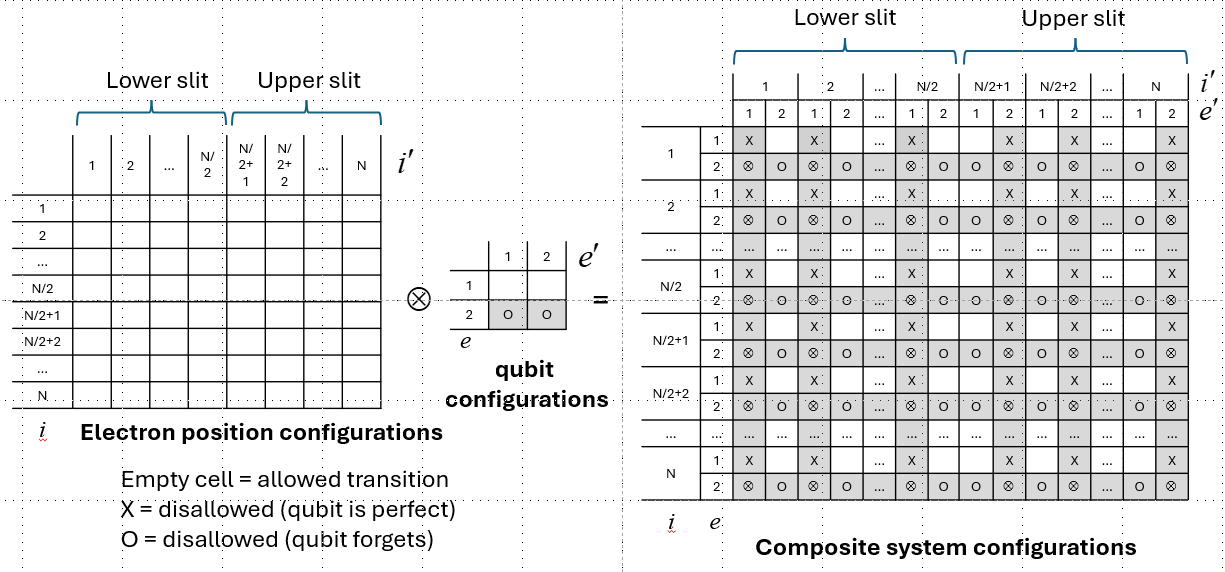}
	\caption{Transition matrix for composite system-qubit configuration states when qubit detects but then \enquote{Forgets} the electron's entry slit.}
	\label{fig:forgets}
\end{figure}

In the ISP paradigm, the Hilbert space representation of quantum states is replaced by a unitary amplitude transition matrix. In principle, such a matrix may have infinite dimension, but for computational and representational purposes we treat the matrix here as finite. The graphical presentation of state transitions in Figures~\ref{fig:inactive}, \ref{fig:remembers}, and \ref{fig:forgets} may provide insight into the effect of qubit behavior on interference.

The $N\times N$ electron position configurations matrix on the left in these Figures represents a state transition matrix of position $x'_{i'}$ from the slit wall to position $x_i$ on the detector screen. Slit wall and detector screen states correspond to columns and rows, respectively. The empty element cells indicate that all $N\times N$ transitions are in principle allowed. The $2\times 2$ qubit configurations matrix in these Figures, however, has structural constraints that disallow certain transitions depending on qubit behavior. Finally the tensor product composite system configurations matrix on the right adds additional disallowed transitions that set entire columns to zero depending on the interaction (or not) of the qubit with the electron entry slit. Disallowed transition elements are grayed out in these Figures and are set to zero. The reason for omitting these transitions is either due to qubit behavior near time $t'$ (qubit is inactive or perfectly reliable noted with a $\times$) or behavior between times $t'$ and $t$ (qubit remembers or forgets noted with a $\circ$).

Figure~\ref{fig:inactive} portrays a composite matrix in which the qubit is inactive (essentially not present) and remains constantly in state 1, not interacting or entangled with the electron's entry slit. In this \enquote{None} behavior, it is not even necessary to consider the qubit transition matrix, but it is included here for comparison with the \enquote{Remembers} and \enquote{Forgets} behaviors described below. Notice that qubit states $e'=e=2$ are disallowed because the qubit state always remains at its default state $e'=e=1$. The resulting geometric pattern of grayed-out cells in the qubit configuration matrix is repeated in the tensor product matrix. Importantly, the resulting pattern is such that the $N$ allowed screen configurations (with $e=1$) each receives amplitude contributions from both slits, allowing for potential interference.

Figure~\ref{fig:remembers} portrays the situation in which an active qubit reliably detects (just before $t'$) and \enquote{Remembers} (from $t'$ to $t$) the electron entry slit. Notice that because the qubit states at times $t'$ and $t$ must agree, the diagonal qubit configurations that disagree are disallowed. This grayed-out pattern is reflected in the tensor product matrix. Further, because the qubit state at $t'$ must reflect the electron entry slit, lower slit $e'=2$ and upper slit $e'=1$ columns of the tensor product matrix are disallowed. Importantly, the resulting pattern in the tensor product matrix is such that, while all $2N$ screen configurations receive amplitude contributions, none receives contributions from both slits. Thus interference should not be expected.

Figure~\ref{fig:forgets} portrays the situation in which the qubit faithfully records the electron entry slit just before time $t'$, however during the transition between times $t'$ and $t$, it \enquote{Forgets} that information and reverts back to its default state of 1. This places different constraints on the allowed transitions for the composite matrix. Again we note that only $N$ of the $2N\times 2$ transitions are possible for each screen position $x_i$. However, this time, each of the $2N$ screen quantum states at time $t$ receive amplitude contributions from both slits, allowing for the possibility of interference.

It is notable from the tensor product matrices in Figures~\ref{fig:inactive}, \ref{fig:remembers}, and \ref{fig:forgets} that only 1 of the 2 possible qubit states in each $x'_{i'}$ slit position actively contributes amplitude to the screen, regardless of the qubit behavior. Thus the same discrete uniform amplitude distribution across slit positions applies in all 3~cases.

\begin{figure}[htbp]
	\centering
	\includegraphics[width=0.9\linewidth]{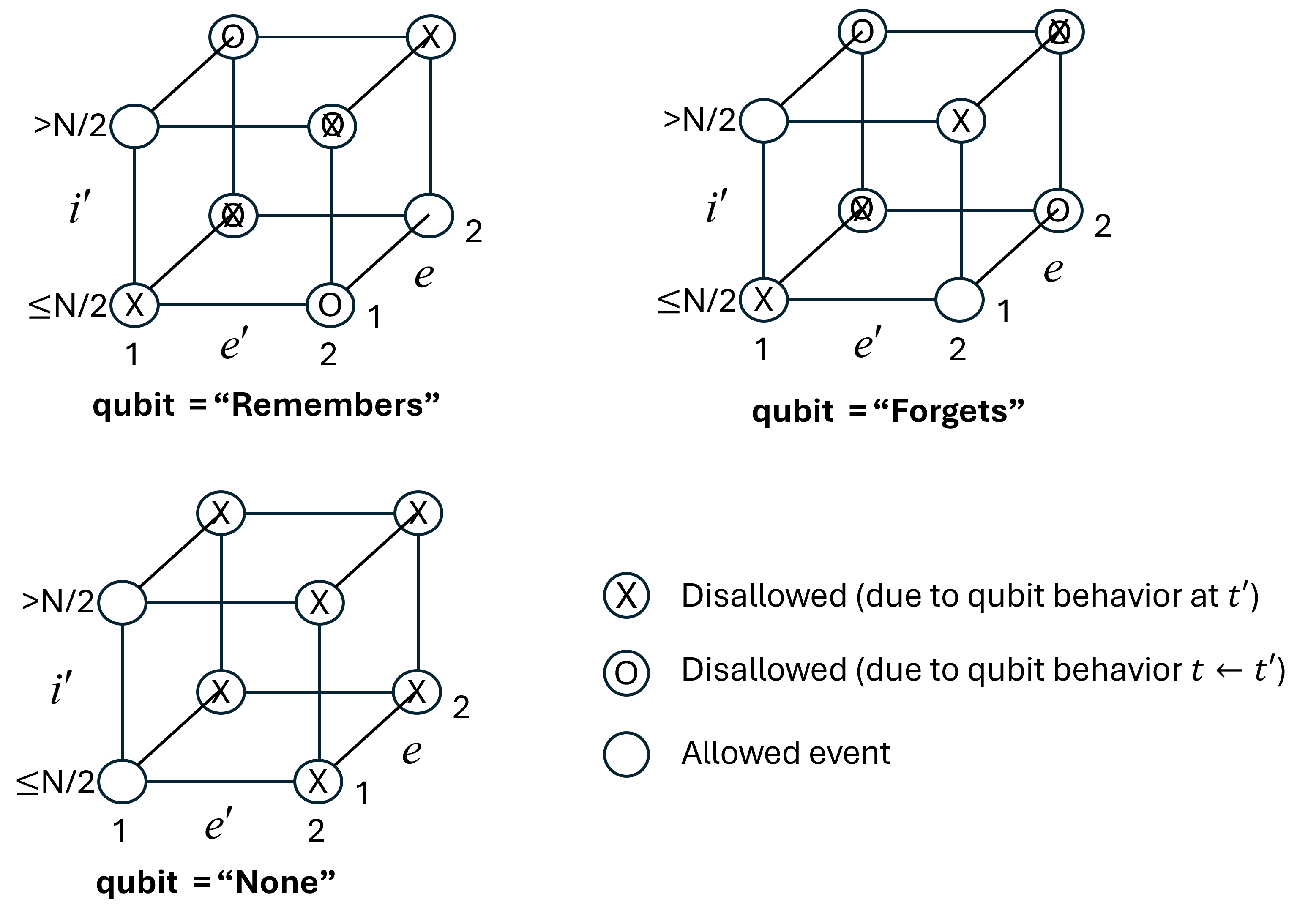}
	\caption{Graphic displaying the active and inactive states and transitions as a function of qubit behavior. Note: $i'\le N/2\equiv$ lower slit and $i'> N/2\equiv$ upper slit.}
	\label{fig:qubit}
\end{figure}

Figure~\ref{fig:qubit} provides an alternative graphic that offers some intuition about the behavior of the composite transition matrix. The influence of the qubit depends only on the indices $i'$, $e'$, and $e$. This allows a 3D cubical plot of these indices for each qubit behavior type. Each of the 8 possible value combinations is classified as allowed or disallowed (and why). 

The pattern of allowed and disallowed combinations in Figure~\ref{fig:qubit} are based on the behavioral rules:

\begin{itemize}
\item \enquote{None}: $e=e'=1$ (qubit state always remains at its default value of 1)

\item \enquote{Remembers}: $e'$ must always be consistent with $i'$ (since qubit is perfectly reliable) and $e=e'$ since qubit "Remembers"

\item \enquote{Forgets}: $e'$ must always be consistent with $i'$ (since qubit is perfectly reliable) and $e=1$ (since qubit always reverts to the default value of 1)
\end{itemize}

Only 2 value combinations are allowed for any of the 3 behaviors. For the \enquote{None} and \enquote{Forgets} behaviors (which permit interference), the allowed transitions contribute to the same value of $e$, but for the \enquote{Remembers} behavior (which denies interference), the allowed transitions contribute to different values of $e$. 

\section{ISP modeling of the double slit experiment}

In this section we are led by the results provided in \cite{lecturenotes}. Following equations (22.17), (22.33), and (22.36) we identify the most general standalone probability for the electron at time $t$ as given by Equation~\eqref{E:eq1}.

\begin{equation}\label{E:eq1}
	p_i(t) = \left| \psi^{\text{upper}}_{i,1} + \psi^{\text{lower}}_{i,1} \right|^2 + \left| \psi^{\text{upper}}_{i,2} + \psi^{\text{lower}}_{i,2} \right|^2
\end{equation}

Here $\psi^{\text{lower}}_{i,e}$ and $\psi^{\text{upper}}_{i,e}$ represent amplitude contributions from the lower and upper slits respectively at screen positions $x_i$. We recognize upper and lower screen contributions to the $\emph{same}$ value of $i=1,2,...,N$ and $e=1,2$ as inclusive events whose amplitudes are additive, while exclusive events leading to $\emph{different}$ quantum configurations, exhibit independent probability contributions.

We note that Equation~\eqref{E:eq1} implicitly mirrors the Born rule (QM axiom number 4 above). In the ISP paradigm, Equation~\eqref{E:eq1} is based on identifying marginal stochastic probabilities at time $t$ with sums of conditional amplitudes obtained by the action of the elements of time evolution (i.e., $t \leftarrow t'$) operators on the elements of configuration state amplitude vector at time $t'$.  

Equations~\eqref{E:eq2} and \eqref{E:eq3} define these marginal amplitude contributions from lower and upper slits, respectively.

\begin{equation}\label{E:eq2}
	\psi^{\text{lower}}_{i,e} = \sum_{i'=1}^{N/2}\sum_{e'=1}^{2} U^{SE}_{(i,e),(i',e')}(t \leftarrow t') \times \psi_{i',e'}(t')
\end{equation}

\begin{equation}\label{E:eq3}
	\psi^{\text{upper}}_{i,e} = \sum_{i'=N/2+1}^{N}\sum_{e'=1}^{2} U^{SE}_{(i,e),(i',e')}(t \leftarrow t') \times \psi_{i',e'}(t')
\end{equation}

As mentioned earlier, we assume that the probability density distribution over the range of $x'$ within the 2 slits forms a continuous uniform distribution with constant probability density $1/(2a)$ that corresponds to a probability amplitude of $1/\sqrt{2a}$. This continuous amplitude distribution is coarse grained by assigning a discrete amplitude of 

\begin{equation}\label{E:eq4}
	\psi_{i',e'}(t')=\frac{\Delta_{slit}}{\sqrt{2a}}\, ,
\end{equation}

\noindent
to each of $N$ discrete levels, evenly spaced on the $x'$ scale. Recall from Figures~\ref{fig:inactive},  \ref{fig:remembers}, and \ref{fig:forgets} that while there are $2N$ unique configurations at time $t'$, only $N$ configurations contribute amplitude to time $t$ configurations, regardless of qubit behavior. 

The coarse-grained width of each $x'_{i'}$ interval is given in Equation~\eqref{E:eq5}.

\begin{equation}\label{E:eq5}
	\Delta_{slit} =\frac{2a}{N}
\end{equation}

Equations~\eqref{E:eq6} to \eqref{E:eq8}  describe the path-integral kernel for a free particle (\cite{feynman}, eq 3.3, page 42). Equation~\eqref{E:eq6} identifies this kernel with the elements of the composite system's time-evolution operator as defined in \cite{lecturenotes}, which also anticipates the exponential form of the operator elements,

\begin{equation}\label{E:eq6}
	U^{SE}_{(i,e),(i',e')}(t \leftarrow t')=Ae^B \, .
\end{equation}
\noindent
Where,

\begin{equation}\label{E:eq7}
	A=\sqrt{\frac{m}{2i\pi\hbar L/v}} \, ,
\end{equation}
\noindent
and

\begin{equation}\label{E:eq8}
	B=\frac{im(x_i-x'_{i'})^2}{2\hbar L/v} \, .
\end{equation}

Equation~\eqref{E:eq9} defines the discrete values of $x_i, i = 1,2,...,N$ at the screen at time $t$, i.e.,

\begin{equation}\label{E:eq9}
	x_i=(i-0.5)\times \Delta_{screen} + Z_{min} \, ,
\end{equation}
\noindent
where the distance between $x_i$ values at the screen is given in Equation~\eqref{E:eq10}.

\begin{equation}\label{E:eq10}
	\Delta_{screen}=\frac{Z_{max}-Z_{min}}{N}
\end{equation}

Equation~\eqref{E:eq11} defines the discrete values of $x'_{i'}, i' = 1,2,...,N$ at the slit wall at time $t'$.

\begin{equation}\label{E:eq11}
	x'_{i'} = (i' - 0.5)\times \Delta_{slit} +
	\begin{cases}
		-\frac{d+a}{2}, & \text{if } i' \le \frac{N}{2} \\
		\frac{d-a}{2}, & \text{if } i' > \frac{N}{2}
	\end{cases}
\end{equation}

\section{Results}
\begin{figure}[htbp]
	\centering
	\includegraphics[width=0.9\linewidth]{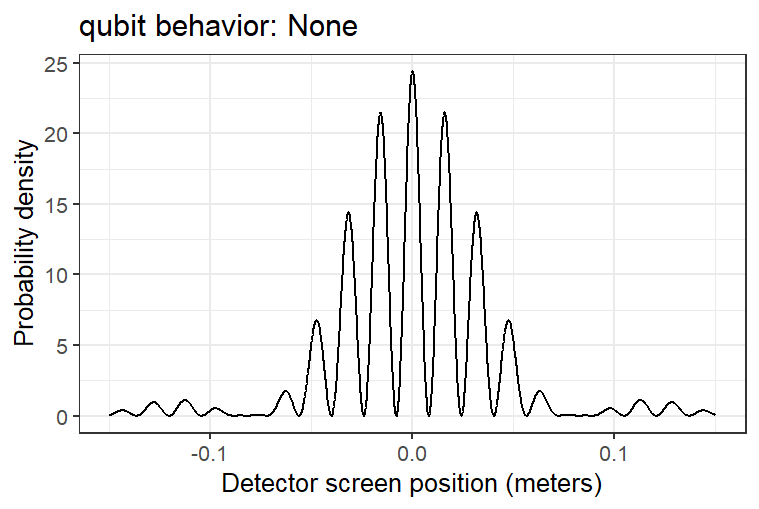}
	\caption{When the qubit is innactive, interference fringes are evident.}
	\label{fig:nonep}
\end{figure}

\begin{figure}[htbp]
	\centering
	\includegraphics[width=0.9\linewidth]{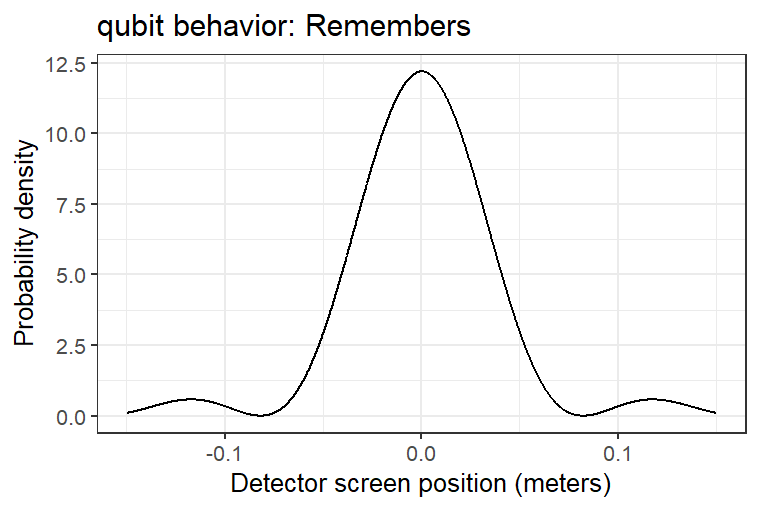}
	\caption{When the qubit observes and records the electron's entry slit, fringes disappear leaving only a diffraction pattern.}
	\label{fig:remembersp}
\end{figure}

\begin{figure}[htbp]
	\centering
	\includegraphics[width=0.9\linewidth]{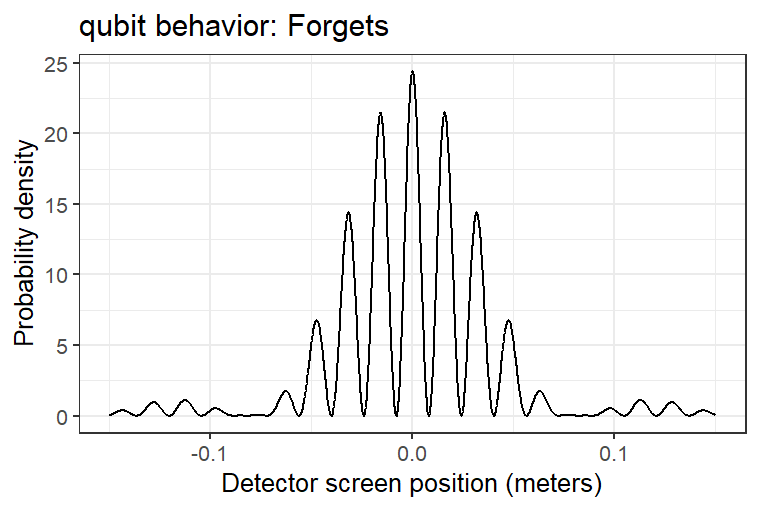}
	\caption{When the qubit observes but forgets the electron's entry slit, fringes re-appear.}
	\label{fig:forgetsp}
\end{figure}

Figures~\ref{fig:nonep}, \ref{fig:remembersp}, and \ref{fig:forgetsp} display the probability density (or intensity) as a function of detector screen position for the \enquote{None}, \enquote{Remembers}, and \enquote{Forgets} qubit behaviors, respectively.  

As anticipated in \cite{lecturenotes}, with the qubit inactive (qubit = \enquote{None}), interference fringes are clearly seen. These fringes disappear when the qubit is actively observing and recording (qubit = \enquote{Remembers}) the slit through which the electron passes. Only the (superimposed) single slit diffraction pattern remains. When the qubit fails to record the slit through which the electron passes (qubit = \enquote{Forgets}), the interference fringes return.

This behavior, which appears surprising and unintuitive when viewed from the foundation of QM axioms, is readily understandable from an ISP perspective. The system $\otimes$ qubit tensor product generates a new set of composite configuration states whose transition behavior follows directly from familiar laws of probability. 

Following the standard derivations of single-slit diffraction and double-slit interference in optics (\cite{hecht}, section 10.2, pages 342-343 and Section 10.3, pages 352-353, respectively), the first diffraction minima occur when $a\sin{\theta}=\pm \lambda$ giving minima at $\approx \pm \lambda L/a = \SI{\pm 0.082}{\meter}$ and the first secondary diffraction maxima occur at approximately $\pm 1.43\lambda L/a = \SI{\pm 0.1173}{\meter}$ from the center. The expected double slit fringe separation distance is $\approx\lambda L/d = \SI{0.02}{\meter}$. These expected values agree well with those observed in Figures~\ref{fig:nonep}, \ref{fig:remembersp}, and \ref{fig:forgetsp}.

As a normalization check on probability (Intensity) at the screen, we find that

$$
\sum_{i=1}^{N}p_i(t)\Delta_{screen} \approx 0.95
$$
regardless of the qubit behavior. Thus the probability mass within the displayed support ($-0.15 < x < +0.15$) in Figures~\ref{fig:nonep}, \ref{fig:remembersp}, and \ref{fig:forgetsp} represents about 95\% of the total. Approximately 5\% of the probability mass lies outside this range.

\section{Discussion and Future Work}
The ISP perspective that is graphically illustrated in Figures~\ref{fig:inactive}, \ref{fig:remembers}, and \ref{fig:forgets} provides an intuitive understanding of the corresponding patterns seen in Figures~\ref{fig:nonep}, \ref{fig:remembersp}, and \ref{fig:forgetsp}. This perspective thus offers the promise of rendering arcane behaviors of quantum systems intuitive and understandable based on familiar arguments from probability theory. Computation is an insightful way to explore more complex features based on the ISP paradigm.

It would be valuable to explore the broader applicability of ISP models. For instance, the present work remains incomplete in that it relies on the path-integral kernel for a free particle (Equations~\eqref{E:eq6}–\eqref{E:eq8}). Developing an ISP formulation that independently reproduces the path-integral result would provide a more fundamental and self-contained demonstration of the framework’s explanatory power.

\begingroup
\sloppy
\bibliography{references}
\endgroup

\section{Appendix}

 The R function \texttt{condition} shown in Listing~\ref{lst:qubitcode} implements the qubit behavior described above and illustrated in Figure~\ref{fig:qubit}. The function operates on the variable \texttt{qubit} based on the input values of \texttt{N}, \texttt{ip} ($i'$, which indexes the electrons position at the wall), \texttt{ep} ($e'$, which indexes the qubit state at the wall), and \texttt{e\_} ($e$, which indexes the qubit state at the screen). The function outputs the logical variable \texttt{include}, which affects the value of the $[(i,i'),(e,e')]^{\text{th}}$ element of the transitional amplitude composite system–environment matrix $U^{SE}_{(i,e),(i',e')}(t \leftarrow t') \times \psi_{i',e'}(t')$ (see Equations~\eqref{E:eq2} and \eqref{E:eq3} in the text and the R code in Listing~\ref{lst:appcode}).
 
 {\footnotesize
 	\begin{lstlisting}[language=R, caption={Function that encodes the behavior of the qubit}, label={lst:qubitcode}]
 		condition<-function(qubit,N,ip,ep,e_){
 			include<-FALSE # default value
 			if(qubit=="None" & ep==1 & e_==1){ 
 				include<-TRUE}
 			if(qubit=="Remembers" &
 			((ip<=N/2 & ep==2)|(ip>N/2 & ep==1)) &
 			(ep==e_)){include<-TRUE}
 			if(qubit=="Forgets" &
 			((ip<=N/2 & ep==2)|(ip>N/2 & ep==1)) & 
 			(e_==1)){include<-TRUE}
 			include}
 	\end{lstlisting}
 }
 
 The R code in Listing~\ref{lst:appcode} produces Figures~\ref{fig:inactive}, \ref{fig:remembers}, and \ref{fig:forgets}. The sums of amplitudes from each  of the \texttt{N/2} slit positions (with 2 qubit states per position) are separately accumulated for each slit. The variable \texttt{psi\_screen1} accumulates contributions from slit 1 (upper slit) while \texttt{psi\_screen2} accumulates contributions from slit 2 (lower slit). Variables \texttt{psi\_screen1} and \texttt{psi\_screen2} are each Nx2 matrices (N positions with 2 qubit states per position) initialized with zeros. The accumulated values depend on the value of \texttt{include} as follows:
 
  \begin{itemize}
 	\item If \texttt{include = FALSE}, the value is set to zero (disallowed transition).
 	\item If \texttt{include = TRUE}, the value is set to that of the respective element of the transitional amplitude composite system–environment matrix (allowed transition).
 \end{itemize}
 
 This code approximates the continuous probability amplitude distribution at the screen as a discrete distribution with \texttt{N} levels. To account for this and assure normalization, the factor \texttt{Delta\_slit1} (or \texttt{Delta\_slit2}) is included in the calculation of \texttt{U\_i\_e\_ip\_ep}.
 
{\footnotesize
\begin{lstlisting}[language=R, caption={ R code used for computations}, label={lst:appcode}]
rm(list = ls()) # clear global environment
library(tidyverse)
# Description of the electron
 lamda <- 0.0123*10^(-8)
 h <- 6.62607015*10^(-34)
 hbar <- h/(2*pi) 
 me <- 9.109*10^-(31) 
 p <- h/lamda #momentum (kg*meters/sec)
 v <-  p/me # velocity (meters/sec)
 N<-2000 
# Description of the slit wall
 a <- 0.15*10^(-8) 
 psi_slit <- 1/sqrt(2*a)
 d <- 0.615*10^(-8) 
 Zmax_slit1 <-   (d + a)/2
 Zmin_slit1 <-   (d - a)/2
 Delta_slit1 <- (Zmax_slit1 - Zmin_slit1)/(N/2)
 Zmax_slit2 <- - (d - a)/2
 Zmin_slit2 <- - (d + a)/2
 Delta_slit2 <- (Zmax_slit2 - Zmin_slit2)/(N/2)
# Description of the screen
 L <- 1 
 Zmin_screen <- -0.15
 Zmax_screen <- +0.15
 Delta_screen <- (Zmax_screen - Zmin_screen)/N
# initialize
 psi_screen1 <- matrix(0,nrow=N,ncol=2)
 psi_screen2 <- matrix(0,nrow=N,ncol=2)
 qubit <- "None" # or "Remembers" or "Forgets"
# Iterations
for(i in 1:N){ 
 x <- (i-1)*Delta_screen + Delta_screen/2 + Zmin_screen
 for(e_ in 1:2){ 
  # Equation 2
  for(ip in 1:(N/2)){ 
   for(ep in 1:2){ 
	if(condition(qubit,N, ip, ep, e_)){
     w<-(ip-1)*Delta_slit2 + Delta_slit2/2 + Zmin_slit2
     A<- sqrt(me/(2*1i*pi*hbar*L/v))
     B<-1i*me*( ((x-w)^2)/(L/v))/2/hbar
     U_i_e_ip_ep<-  A*exp(B)*psi_slit*Delta_slit2 
     psi_screen2[i,e_] <- psi_screen2[i,e_] + U_i_e_ip_ep
    }
   }
  }
  # Equation 3
  for(ip in (N/2+1):N){ 
   for(ep in 1:2){ 
    if(condition(qubit,N, ip, ep, e_)){
     w<-(ip-1)*Delta_slit1 + Delta_slit1/2 + Zmin_slit1
     A<-sqrt(me/(2*1i*pi*hbar*L/v))
     B<-1i*me*( ((x-w)^2)/(L/v))/2/hbar
     U_i_e_ip_ep<-A*exp(B)*psi_slit*Delta_slit1
     psi_screen1[i,e_] <- psi_screen1[i,e_] + U_i_e_ip_ep
    }
   }
  }
 }
}

# Equation 1
Intensity<-
 Re(Conj(psi_screen1[,1]+psi_screen2[,1])*
        (psi_screen1[,1]+psi_screen2[,1])) +
 Re(Conj(psi_screen1[,2]+psi_screen2[,2])*
        (psi_screen1[,2]+psi_screen2[,2]))

d2 <- tibble(
 x=(1:N)*Delta_screen+Delta_screen/2+Zmin_screen,
 P=Intensity
)

ggplot(d2,aes(x=x, y=P)) +
 geom_line() +
 theme_bw() +
 labs(title=paste("qubit behavior:",qubit),
  y="Probability density", 
  x="Detector screen position (meters)")

# Normalization check at the slit wall
(2*a)*psi_slit^2 # continuous pdf
N*Delta_slit1*psi_slit^2 # discrete pdf

# Normalization check at the screen
sum(Intensity*Delta_screen)

# Expected first diffraction minima
# (single slit) 0.082 meters, either 
# side of central maximum
lamda*L/a 

# Expected position of the first side 
# band (single slit) 0.1173 meters 
# either side of central maximum
1.43*lamda*L/a 

# Expected distance between interference 
# fringes (double slit) 0.02 meters appart
lamda*L/d 
	\end{lstlisting}
}

\clearpage

 For reference, Table~\ref{tab:varmap} below lists the variable names used in this paper and gives the correspondence between names used in the R code calculations and the names used in the text.

\begin{table*}[!h]
	\centering
	\caption{Correspondence between equation variables and R code variables.}
	\label{tab:varmap}
	\renewcommand{\arraystretch}{1.2} % adds a little extra row height
	\begin{tabularx}{\textwidth}{l l X l}
		\toprule
		\textbf{Code name} & \textbf{Text name} & \textbf{Definition} & \textbf{Units} \\
		\midrule
		\texttt{lamda} & $\lambda$ & electron wavelength & $\si{\meter}$ \\
		\texttt{hbar} & $\hbar$ & reduced Plank constant & $\si{\joule\second}$ \\
		\texttt{me} & $m$ & electron mass & kg \\
		\texttt{v} & $v$ & electron velocity & $\si{\meter\per\second}$ \\
		\texttt{N} & $N$ & number of positions in wall/screen & count \\
		\texttt{a} & $a$ & slit width & $\si{\meter}$ \\
		\texttt{d} & $d$ & distance between slit centers & $\si{\meter}$ \\
		\texttt{psi\_slit} & $1/\sqrt{2a}$ & amplitude density at slit (See Eq. (4)) & $\si{\metre^{-1/2}}$ \\ 
		\texttt{Zmax\_slit1} & $(d+a)/2$ & position of upper slit top edge & $\si{\meter}$ \\
		\texttt{Zmin\_slit1} & $(d-a)/2$ & position of upper slit bottom edge & $\si{\meter}$ \\
		\texttt{Zmax\_slit2} & $-(d-a)/2$ & position of lower slit top edge & $\si{\meter}$ \\
		\texttt{Zmin\_slit2} & $-(d+a)/2$ & position of lower slit bottom edge & $\si{\meter}$ \\
		\texttt{Delta\_slit1(2)} & $\Delta_{slit}$ & slit interval (see Eq. (5), common to both slits) & $\si{\meter}$ \\
		\texttt{ip} & $i'$ & wall position index & index \\
		\texttt{ep} & $e'$ & wall quibit state index & index \\
		\texttt{w} &  $x'_{i'}$ &  $i'^{th}$ position on wall at time $t'$ (see Eq. (11)) & $\si{\meter}$ \\
		\texttt{L} & $L$ & wall to screen distance & $\si{\meter}$ \\
		\texttt{Zmin\_screen} & $Z_{min}$ &  position of screen bottom & $\si{\meter}$ \\
		\texttt{Zmax\_screen} & $Z_{max}$ &  position of screen top & $\si{\meter}$ \\ 
		\texttt{Delta\_screen} & $\Delta_{screen}$ & screen interval (see Eq. (10)) & $\si{\meter}$ \\       
		\texttt{i} & $i$ & screen position index & index \\
		\texttt{e\_} & $e$ & screen quibit state index & index \\        
		\texttt{x} & $x_i$ & position on screen (see Eq. (9)) & $\si{\meter}$ \\
		\texttt{A} & $A$ & normalizing factor (see Eq. (7)) & $\si{\per\meter}$ \\
		\texttt{B} & $B$ & phase factor (see Eq. (8)) & unitless \\
		\texttt{psi\_screen1} & $\psi^{\text{upper}}_{i,e}$ & $N\times2$ amplitude density from lower slit at time $t$ & $\si{\metre^{-1/2}}$ \\
		\texttt{psi\_screen2} & $\psi^{\text{lower}}_{i,e}$ & $N\times2$ amplitude density from upper slit at time $t$ & $\si{\metre^{-1/2}}$ \\
		\texttt{U\_i\_e\_ip\_ep} & $U^{SE}_{(i,e),(i',e')}\psi_{i',e'}$ & amplitude transition element (see Eq. (6)) & $\si{\metre^{-1/2}}$ \\
		\texttt{Intensity} & $p_i(t)$ & standalone probability  at $(x_i,t)$ & unitless \\
		\bottomrule
	\end{tabularx}
\end{table*}

\section*{Acknowledgments}
\begin{center}
\begin{minipage}{0.8\textwidth}  % Adjust width (0.7–0.9 looks nice)
The author gratefully acknowledges Professor Jacob A. Barandes of Harvard University for his generous instruction and encouragement, and Diane Wolden for her careful technical review of the manuscript.
\end{minipage}
\end{center}
	
\bigskip
\hrule
\bigskip

\section*{About the Author}
\begin{center}
	\begin{minipage}{0.8\textwidth}  % Adjust width (0.7–0.9 looks nice)
David LeBlond received his Ph.D. in Biochemistry from Michigan State University, and M.S. in Statistics from Colorado State University. His professional work has focused on statistical methods in biomedical research and applied modeling. His current interests include the foundations of quantum mechanics and interdisciplinary applications of probability theory in physics.
\end{minipage}
\end{center}	
	
\end{document}